\documentclass[twocolumn,aps,prl,superscriptaddress]{revtex4-2}

\usepackage{graphicx}
\usepackage[dvipsnames]{xcolor}
\usepackage[colorlinks=true,urlcolor=RoyalBlue,citecolor=RoyalBlue,linkcolor=Black]{hyperref}
\usepackage{amsmath}
\usepackage{amssymb}
\usepackage{xr}
\usepackage{upgreek}

\newcommand{\plab}[1]{\textbf{#1}}

\let\oldcite\cite
\renewcommand{\cite}[1]{\mbox{\oldcite{#1}}}

\renewcommand{\mu}{\upmu}

\begin{document}

\title{Non-Gaussian displacements in active transport on a carpet of motile
cells}

\author{Robert Gro{\ss}mann}
\email[Correspondence should be addressed to: ]{rgrossmann@uni-potsdam.de,
rmetzler@uni-potsdam.de, beta@uni-potsdam.de}
\affiliation{Institute of Physics and Astronomy, University of Potsdam,
Potsdam 14476, Germany}
\author{Lara S. Bort}
\affiliation{Institute of Physics and Astronomy, University of Potsdam,
Potsdam 14476, Germany}
\author{Ted Moldenhawer}
\affiliation{Institute of Physics and Astronomy, University of Potsdam,
Potsdam 14476, Germany}
\author{Setareh Sharifi Panah}
\affiliation{Institute of Physics and Astronomy, University of Potsdam,
Potsdam 14476, Germany}
\author{Ralf Metzler}
\affiliation{Institute of Physics and Astronomy, University of Potsdam,
Potsdam 14476, Germany}
\affiliation{Asia Pacific Center for Theoretical Physics, Pohang 37673,
Republic of Korea}
\author{Carsten Beta}
\affiliation{Institute of Physics and Astronomy, University of Potsdam,
Potsdam 14476, Germany}

\date{\today}
	
\begin{abstract}	
We study the dynamics of micron-sized particles on a layer of motile cells.
This cell carpet acts as an active bath that propels passive tracer 
particles via direct mechanical contact. The resulting nonequilibrium transport
shows a crossover from superdiffusive to normal-diffusive dynamics. The
particle displacement distribution is distinctly non-Gaussian even
in the limit of long measurement times---different from typically reported
Fickian yet non-Gaussian transport, for which Gaussianity is restored
beyond some system-specific correlation time. We obtain the distribution
of diffusion coefficients from the experimental data and introduce a model
for the displacement distribution that matches the experimentally observed
non-Gaussian statistics and argue why similar transport properties are expected for many
composite active matter systems.
\end{abstract}

\maketitle

The collective behavior of active particles is the focus of
one of the most dynamically evolving research directions in nonequilibrium
statistical and biological physics over the past decade. Active 
particles~\cite{romanczuk_active_2012,bechinger_active_2016} convert chemical energy
into motion and provide a unifying concept for a wide range of systems, such
as engineered bipolar `Janus particles' with different sources of 
activity~\cite{soto_self_2014,feldmann_manipulation_2016,feldmann_2019_light,zheng_2013_nongauss}, 
bacterial swimming~\cite{berg1993random,berg_ecoli_2004} and swarming~\cite{beer_statistical_2019}, 
crawling cells~\cite{aranson_physical_2016}, bristle robots~(hexbugs)~\cite{altshuler_2023_environmental,dauchot_2019_dynamics,horvath_2023_bouncing}, or groups of foraging animals~\cite{viswanathan_physics_2011}.
Large ensembles of interacting active particles, in which energy is
continuously injected and dissipated locally, operate far from thermodynamic
equilibrium. This gives rise to a plethora of nonequilibrium phenomena, such
as the emergence of large scale patterns~\cite{marchetti_hydrodynamics_2013},
topological order~\cite{shankar_topological_2022}, or nonequilibrium phase
separations~\cite{cates_motility_2015,chate_dadam_2020,baer_self_2020}, and
raises questions about the thermodynamics of such systems, e.g., their 
pressure~\cite{solon_pressure_2015,ginot_2015_nonequilibrium,nikola_2016_active}. All these features are being studied under the
common theme of active matter physics~\cite{hallatschek_2023_proliferating,bowick_2022_symmetry}.

In many real-world settings, active agents interact with passive objects
in their surroundings that introduce additional degrees of complexity
(`composite active matter'~\cite{lepro_optimal_2022}). For instance, boundaries
and obstacles may rectify their motion~\cite{lambert_collective_2010}, self-similar structures may emerge
at interfaces in active matter invasion~\cite{xu_2023_geometrical}, or non-isotropic
passive objects, such as gears~\cite{Sokolov2009,ray_rectified_2023}
or curved tracers~\cite{mallory_curvature_2014} can be powered by an
`active bath' of self-propelled particles to perform coherent motion. For a fundamental understanding as well as for many
applications, the statistics of transport in a bath of active elements is
of particular importance. Earlier work has focused on tracer diffusion in
active fluids composed of suspensions of biological swimmers,
such as bacteria or algae, agitating the surrounding fluid of the
bath~\cite{wu_particle_2000,mino_induced_2013,leptos_dynamics_2009},
or particles in the vicinity of flow-generating active carpets~\cite{guzman_active_2021}. 
Non-trivial scalings with several crossovers
in the mean-squared displacement~(MSD) of passive tracer particles were
observed; for a review, see Ref.~\cite{bechinger_active_2016}. However, little
is known about the statistical properties of other types of active baths. A
particularly large and important class are composite systems, in which the
interactions between active elements and passive tracers are established
by direct mechanical contact and adhesion instead of fluid flows and
hydrodynamic interactions. This situation arises, for example, when slowly
moving, adherent cells interact with passive objects, and it has important
practical implications for the movement of foreign bodies in tissues or the
delivery of drug-loaded particles in a multicellular environment.

\begin{figure*}
\includegraphics[width=\textwidth]{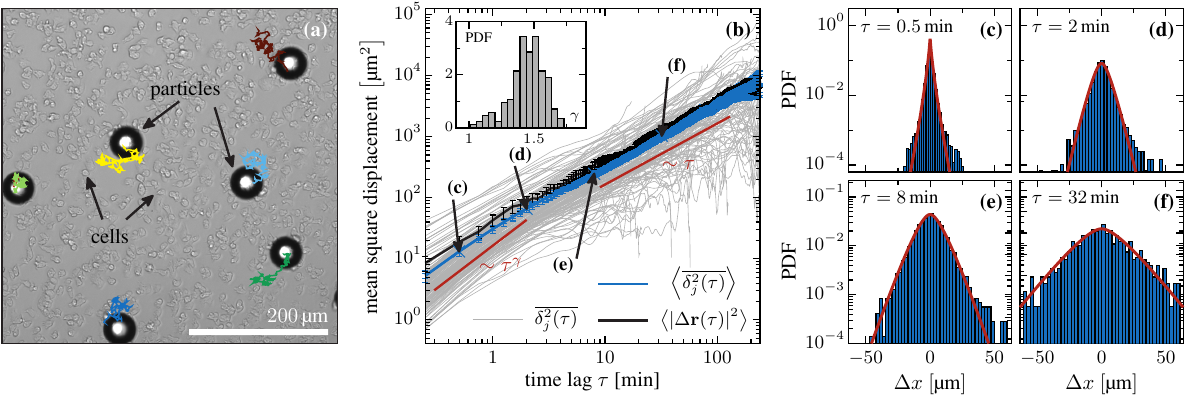}
\vspace{-0.6cm}
\caption{Characteristics of cargo particle transport on cellular
monolayers. \plab{(a)} Illustration of experimental bright-field
microscopy recording with particle trajectories displayed as colored
overlays~(see SM~\cite{noteSI} for a movie). Bright spots with a black halo are particle~($46\,\mu\mathrm{m}$ diameter). Cells with an extension of about
$10\,\mu\mathrm{m}$ appear translucent in the background. 
\plab{(b)} MSD as function of time: TAMSDs $\overline{\delta_j^2(\tau)}$, Eq.~\eqref{eqn:TAMSD}, of individual particles~(light grey lines), their ensemble average $\langle\overline{\delta^2_j(\tau)}\rangle$ (blue) and the
ensemble-averaged MSD $\langle|\Delta\mathbf{r}\!\left(\tau\right)\!|^2\rangle$ (black). 
Error bars indicate $1\sigma$-confidence intervals. These
MSDs reveal two distinct regimes: superdiffusion at short time scales and
normal diffusion at long times. The inset shows the scaling exponents~$\gamma$ 
obtained by fitting~$\overline{\delta_j^2(\tau)} \sim \tau^{\gamma}$ to the first three
data points of TAMSDs of individual particles~(time interval:~$45 \, \mbox{s}$). 
The interaction of a cargo with a single cell may induce anomalous
scaling---in line with experimental recordings of individual cell-cargo
pairs~\cite{lepro_optimal_2022}---whereas the collective long-time particle
transport by several cells is normal. Panels \plab{(c)}-\plab{(f)} show
the displacement PDFs along the $x$-axis for different lag times. These PDFs
reveal distinctly non-Gaussian statistics, even for long lag times. For short
lag times, the displacement PDF is well-fitted by a stretched exponential
distribution, proportional to $\exp (-a|\Delta x|^\delta)$, with an exponent
$\delta \approx 0.77$~(red line in~\plab{(c)}). The red
lines in panels \plab{(d)}-\plab{(f)} depict the predicted displacement
distributions, based on heterogeneous Brownian motion [cf.~Eqs.~\eqref{eqn:prop_prdc_t} and main text for details].}
\label{fig:msd_dist_overview}
\end{figure*}

Here we consider such a composite biohybrid system as a paradigmatic showcase
of an active bath, in which passive objects are agitated and transported
by self-propelled agents via direct mechanical contact. As an active
bath, we use a monolayer of cells of the social amoeboid {\it Dictyostelium
discoideum}, an established model organism with well-characterized properties~\cite{annesley_dicty_2009}. 
As {\it D.~discoideum} cells show 
unspecific adhesion to most common material surfaces~\cite{loomis_innate_2012},
adhesive contacts between cells and microparticles are formed upon collision
and may spontaneously break again~\cite{Nagel2018}. No specific surface
functionalization is required. 
Unbound cells can freely move over the two-dimensional substrate. 
When cells bind to the microparticles,
the active motion of cells results in nonthermal fluctuating forces
that randomly displace the particles. While this process
has been studied in detail for single cells interacting with a single
particle~\cite{lepro_optimal_2022,panah_cargo_2023}, we here consider
particles that are attached to many cells at the same time. We performed
time-lapse recordings with a time interval of $15\,\mbox{s}$ between frames
over a duration of 4 hours. In total, 174 particle trajectories were
extracted; the majority  were longer than 2.5 hours~\cite{noteSI}. 
An example from 
the recorded image stacks is shown in Fig.~\ref{fig:msd_dist_overview}(a),
with the particle trajectories displayed as colored overlays~(see Material and
Methods in SM~\cite{noteSI} for experimental details). From the trajectories,
we characterized the dynamics of particles on the cell layer in terms
of their MSD, their displacement probability density functions~(PDFs), and
their displacement auto-correlation function~(DACF).

\emph{MSD crossover from superdiffusive spreading to normal diffusion.}
Position time traces $\mathbf{r}_j(t)$ of length $T$ of an individual particle
$j$ are evaluated in terms of the time-averaged MSD~(TAMSD)~\cite{barkai_2012_strange}
\begin{eqnarray}
\label{eqn:TAMSD}
\overline{\delta^2_j(\tau)}=\frac{1}{T-\tau}\int_0^{T-\tau} \left[\mathbf{r}_j
(t+\tau)-\mathbf{r}_j(t)\right]^2 dt,
\end{eqnarray}
where~$\tau$ is the lag time. The ensemble mean-TAMSD is defined
as $\langle\overline{\delta^2_j(\tau)}\rangle=N^{-1}\sum_{j=1}^N\overline{\delta
^2_j(\tau)}$. From an ensemble of particles one can also determine the
ensemble-averaged MSD $\langle\Delta\mathbf{r}(\tau)^2\rangle=N^{-1}\sum_{j=1}^N
[\mathbf{r}_j(\tau)-\mathbf{r}_j(0)]^2$. 
In an ergodic system the TAMSD converges to the ensemble-averaged MSD in the long time limit~$\tau/T\to0$; otherwise the dynamics is non-ergodic~\cite{he_2008_random,burov_aging_2010,barkai_2012_strange}. 

The results for MSD and TAMSD are displayed in
Fig.~\ref{fig:msd_dist_overview}(b). The TAMSDs~(light grey lines)
reveal a large amplitude spread, indicating significant differences in the transport
of individual particles. Their ensemble average, displayed as the blue line, 
exhibits two regimes: superdiffusion $\langle\Delta\mathbf{r}(\tau)^2\rangle\simeq
\tau^{\gamma}$ at short lag times with an anomalous diffusion exponent of
approximately $\gamma\approx1.45$~(median) and normal diffusion~($\gamma\approx1$) at
long lag times. Before the crossover time $\tau\approx2\,\mbox{min}$, individual
TAMSD scaling exponents vary between $\gamma\approx1.33$ and $1.57$~($1\sigma$-interval); 
the inset in Fig.~\ref{fig:msd_dist_overview}(b) shows a histogram of the exponents~$\gamma$ 
obtained from fits to TAMSDs. 

A similar superdiffusive scaling was observed for the MSD of single
cell trajectories~\cite{dieterich_anomalous_2008,li_persistent_2008,takagi_functional_2008,
makarava_quantifying_2014,Cherstvy2018} and is reflected in the trajectories of
cargo particles transported by an individual cell~\cite{lepro_optimal_2022}. The
crossover time to normal diffusion corresponds to a length scale that is comparable
to the average cell size~($5$ to $10\,\mu\mathrm{m}$ in radius). We thus
conclude that the short-time superdiffusive scaling reflects the action of
individual cells, while the long-term normal diffusion corresponds to collective
particle transport involving many cells. 

\emph{Non-Gaussian displacement distributions.} The PDFs of the particle
displacements $\Delta\mathbf{r}_j(t,\tau)=\mathbf{r}_j(t+\tau)-\mathbf{r}_j(t)$
are shown in Figs.~\ref{fig:msd_dist_overview}(c-f) for different lag times~$\tau$.
The lag times were chosen from the superdiffusive regime
(Fig.~\ref{fig:msd_dist_overview}(c), $\tau=0.5\,\mbox{min}$), close to the
crossover time~(Fig.~\ref{fig:msd_dist_overview}(d), $\tau=2\,\mbox{min}$),
and from the diffusive regime~(Figs.~\ref{fig:msd_dist_overview}(e) and~(f)
with $\tau=8$ and $32\,\mbox{min}$), respectively. Notably, all
displacement PDFs are non-Gaussian, with a positive excess kurtosis implying
a leptokurtic PDF with a more pronounced peak at zero and heavier tails as
compared to a Gaussian distribution. At short lag times, the displacement PDF
is well approximated by a stretched exponential, proportional to $\exp(-a|
\Delta x|^\delta)$ with $\delta\approx0.77$. With increasing lag time, the
PDF changes to a non-Gaussian shape with exponential tails~($\delta\approx1$)
in the Fickian regime~($\gamma = 1$). Below, we describe this exponential PDF by a
heterogeneous Brownian diffusion model. We also invoke tempered fractional
Laplace motion as a model for all lag times.

\begin{figure}
\includegraphics[width=\columnwidth]{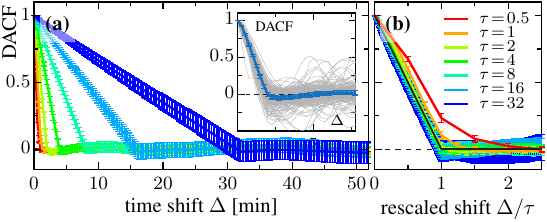}
\vspace{-0.6cm}
\caption{\plab{(a)}~Ensemble-averaged DACFs for different lag times~$\tau$ as listed
in the right panel~(all times in minutes). In the inset, grey lines show
single-particle DACFs $\tilde{C}_{\tau}^{(j)}(\Delta)$ for lag time~$\tau
=2\,\mbox{min}$. The ensemble average $\langle\tilde{C}_{\tau}(\Delta)\rangle$ is
depicted in blue---error bars indicate $3\sigma$-confidence intervals. Panel~\plab{(b)} 
shows the same data as in panel~\plab{(a)}, as function of the rescaled time shift
$\Delta/\tau$. For lag times $\tau\ge2\,\mbox{min}$, the DACF collapses onto a
triangular correlation function~(black solid line), in line with 
Brownian motion~\cite{noteSI}. }
\label{fig:incr_corr}
\end{figure}

\emph{DACF indicates Brownian motion.} The ACF of the displacements $\Delta
\mathbf{r}_j(\tau)$ of particle $j$ is defined as
\begin{align}
\label{eqn:DACF_def}
&C^{(j)}_{\tau}(\Delta)=\overline{\Delta\mathbf{r}_j(t+\Delta,\tau)\cdot\Delta
\mathbf{r}_j(t,\tau)}\\
&=\frac{1}{T-\Delta-\tau}\int_0^{T-\Delta-\tau}\Delta\mathbf{r}_j(t+\Delta,\tau)
\cdot\Delta\mathbf{r}_j(t,\tau)dt\nonumber
\end{align}
and measures the degree of correlation between a particle displacement $\Delta
\mathbf{r}_j(t,\tau)$ in a time interval~$\tau$ starting at time $t$ and a
displacement in an interval of the same length~$\tau$, beginning at the later
time $t+\Delta$. For small
time shifts $\Delta$, the displacements are highly correlated, whereas the
correlations decay to zero at longer $\Delta$. The value of the DACF at
$\Delta=0$ is identical to the TAMSD $\overline{\delta^2_j(\tau)}$,
cf.~Eq.~\eqref{eqn:TAMSD}. 
We focus on the $\Delta$-dependence of the normalized DACF $\tilde{C}_{\tau} ^{(j)}(\Delta)=C_{\tau}^{(j)}(\Delta)/C_{\tau}^{(j)}(0)$ as well as on its ensemble mean $\langle\tilde{C}_{\tau}(\Delta)\rangle=N^{-1}\sum_{j=1}^N \tilde{C}_{\tau}^{(j)}(\Delta)$. 
The temporal decay of the ensemble-averaged
DACF is shown in Fig.~\ref{fig:incr_corr}(a) for different time lags~$\tau$;
the DACFs of individual particle trajectories, together with their ensemble
average, are displayed in the inset for $\tau=2\,\mbox{min}$ as an example.
As expected, the degree of correlation increases with~$\tau$. Notably, the
correlations decrease linearly as a function of $\Delta$ for time lags~$\tau$
in the diffusive regime and are essentially zero for time shifts $\Delta\ge
\tau$. This is a signature of independent steps in normal Brownian motion,
for which the renormalized DACF takes the triangular shape $\tilde{C}_{\tau}
(\Delta)=1-|\Delta|/\tau$ for $0\le|\Delta|\le\tau$ and 0 otherwise~\cite{noteSI}. 
In Fig.~\ref{fig:incr_corr}(b) we show the DACFs as a function of
the rescaled time shift $\Delta/\tau$. Indeed, the data collapse onto a single
master curve for all time lags $\tau\ge2\,\mbox{min}$, underlining that particle
displacements become independent at times of several minutes and beyond.

\emph{Fickian yet non-Gaussian particle transport.} Our analysis above reveals
that polystyrene spheres on a carpet of cells show distinct characteristics of Brownian motion above the crossover time: the
MSD increases linearly and the DACF has a triangular shape. However, the
displacement PDFs are non-Gaussian for all considered lag times. We tested
whether the non-Gaussian statistic arises from non-stationary dynamics of the
system but did not detect any statistically significant changes in the bead
dynamics over time~(see SM~\cite{noteSI}).

Non-Gaussian displacement PDFs together with a linear-in-time MSD were observed
in different stationary
systems, i.a., colloids diffusing along linear
tubes or through entangled actin networks~\cite{wang_anomalous_2009}. Such
observations may arise when the diffusion coefficient of a diffusing
particle follows a stochastic diffusion process itself (`diffusing diffusivity')~\cite{chubynsky_diffusing_2014,lanoiselee_2018_diffusion,sabri_2020_elucidating,yamamoto_2021_universal,sakamoto_2023_heterogeneous}. If the temporal diffusivity variation is slower
than the experimentally relevant time scales, the diffusivity is effectively
constant in time but randomly distributed across the diffusing particles~\cite{chechkin_brownian_2017,sposini_random_2018}. The PDF of diffusivities
across the ensemble introduces an additional level of annealed disorder, also
called `superstatistics'~\cite{beck_superstatistics_2003}. For ensembles
of active particles superstatistical diffusivities and speeds were recently 
analyzed theoretically~\cite{lemaitre_2023_nongaussian,khadem_2021_transport}.

\begin{figure*}
\includegraphics[width=0.8\textwidth]{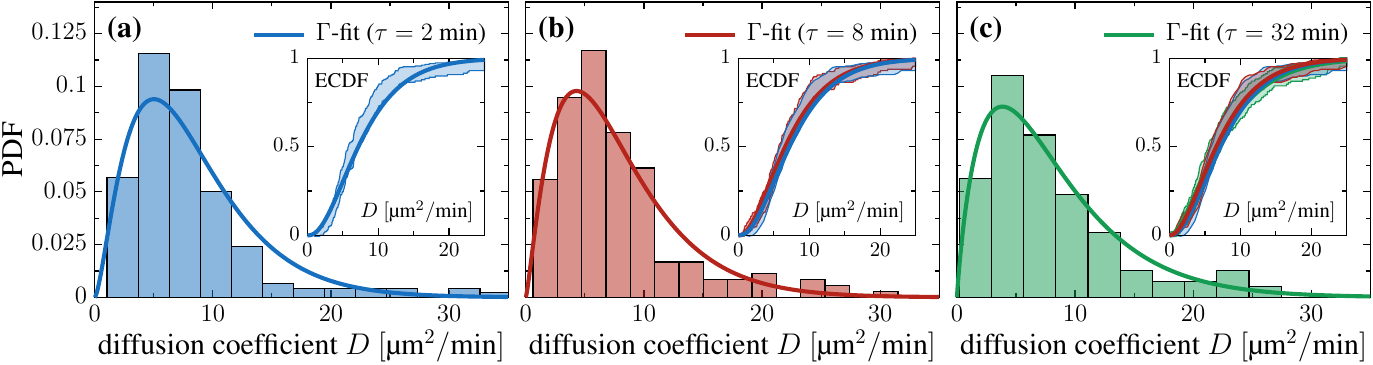}
\vspace{-0.2cm}
\caption{Histograms of diffusion coefficients for different lag times~$\tau$. 
Solid lines: fits with a gamma distribution~[Eq.~(\ref{gamma})], 
inferred by maximum likelihood estimation: \plab{(a)}~$\alpha=2.55$, 
$D_{\beta}=3.23\, \mu\mbox{m}^2/\mbox{min}$, \plab{(b)}
$\alpha=2.26$, $D_{\beta}=3.40\, \mu\mbox{m}^2/\mbox{min}$ and \plab{(c)}
$\alpha=1.91$, $D_{\beta}=4.24\, \mu\mbox{m}^2/\mbox{min}$. Sample means
of diffusion coefficients: \plab{(a)} $\langle D\rangle=(8.2\pm0.5)\,\mu\mathrm{m}^2
/\mbox{min}$ for $\tau=2\,\mbox{min}$, \plab{(b)} $\langle D\rangle=(7.7\pm0.4)\,\mu
\mathrm{m}^2/\mbox{min}$ for $\tau=8\,\mbox{min}$, and \plab{(c)} $\langle D\rangle=(
8.0\pm0.5)\,\mu\mathrm{m}^2/\mbox{min}$ for $\tau=32\,\mbox{min}$. Insets show
the corresponding cumulative distributions:~lines represent fits 
and color-shaded regions are $2\sigma$-confidence intervals of the
empirical distribution function~(ECDF). In the inset of panel \plab{(b)}, the ECDF 
from panel \plab{(a)} is shown as an overlay; the inset of
panel \plab{(c)} contains all three ECDFs. There are 
no statistically significant differences.}
\label{fig:gam_dist_D}
\end{figure*}

\emph{Superstatistics of diffusion coefficients.} Long-time particle motion is Fickian, but individual diffusion coefficients~$D_j$ of particles $j$ vary across the ensemble, resulting in the amplitude
scatter of individual TAMSDs~(grey lines) in Fig.~\ref{fig:msd_dist_overview}(b).
We expect that this variability is caused by variations in the
size and activity within the population of cells~(as resolved on the single cell
level~\cite{Cherstvy2018}). Moreover, the cell density varies in space due to
finite number fluctuations. The spread in $D_j$ may be enhanced by
tug-of-war-style competition between individual cells to which the cargo particle
is attached. 
Even though the statistical properties are identical in the
long-time limit~(independent increments, normal diffusion), quantitative
differences in the diffusivities of individual particles are thus expected.

To test the conjecture that the spread in particle diffusivities may explain the non-Gaussian displacement distributions,
we derive the distribution of diffusivities directly
from experimental data by estimating the $D_j$ values of individual cargo
particles from their TAMSDs via $\hat{D}_j=\overline{\delta_j^2(\tau)}/(4
\tau)$. Fig.~\ref{fig:gam_dist_D} shows histograms of the $D_j$ estimates
for three different lag times~$\tau$. All histograms are nicely fitted by a
gamma PDF
\begin{equation}
\label{gamma}
P(D)= \frac{1}{\Gamma(\alpha) D_{\beta}} \left( \frac{D}{D_{\beta}} \right)^{\! \alpha - 1} \exp \! \left( - \frac{D}{D_{\beta}} \right)
\end{equation}
with shape parameter~$\alpha$ and scale parameter~$D_{\beta}$. 
The goodness of fit can be judged from
the cumulative distributions shown in the insets of Fig.~\ref{fig:gam_dist_D},
where the line corresponds to the fit and the color-shaded region is the $2
\sigma$-confidence interval of the empirical distribution. The gamma
distributions are consistent over time, i.e., the inferred parameter values
are independent of the chosen lag time~$\tau$ and yield a mean diffusion
coefficient of $\langle D\rangle\approx8\,\mu\mathrm{m}^2/\mbox{min}$ in all cases
(see caption of Fig.~\ref{fig:gam_dist_D} for exact numbers). The consistency
of these values supports our conjecture that long-time active transport
of microparticles on a carpet of motile cells is governed by heterogeneous Brownian motion.

\emph{Prediction of displacement PDFs.} We verified the consistency of our
superstatistical model by comparison of the empirical displacement PDFs to
the model prediction. For normal Brownian diffusion, the displacement PDF~(given the diffusion coefficient~$D$) is Gaussian: $\rho(\Delta\mathbf{r},\tau
|D)=(4\pi D\tau)^{-1}\exp[-|\Delta\mathbf{r}|^2/(4D\tau)]$. The displacement
PDF of an ensemble of particles with different diffusivities is then obtained
by averaging with respect to the distribution of diffusion coefficients $P(D)$~\cite{beck_superstatistics_2003,chechkin_brownian_2017}: 
\begin{eqnarray}
\langle\rho(\Delta\mathbf{r},\tau)\rangle=\int_0^{\infty}\rho(\Delta\mathbf{r},
\tau|D)P(D)dD. 
\end{eqnarray}
The corresponding displacement PDFs of the $x$- and $y$-components of $\Delta
\mathbf{r}$ are obtained by marginalization. In the diffusive regime, they were found to be
statistically independent~(linear correlation coefficient below $0.02$ in all
cases). We focus on displacements along the $x$-axis~\footnote{Analogous results are
obtained for displacements along the $y$-axis. }. Using the gamma distribution
model $P(D)$ to describe the heterogeneity in the diffusion coefficients and
the Gaussian propagator $\rho(\Delta\mathbf{r},\tau|D)$, we obtain the
displacement PDF
\begin{subequations}
\label{eqn:prop_prdc_t}
\begin{eqnarray}
\label{eqn:prop_prdc_a}
\!\!\!\!\!\!\!\!\!\!\langle\rho(\Delta x,\tau)\rangle &=&
\mathcal{N}
\frac{|\Delta x|^{\alpha-1/2}}{(D_{\beta}\tau)^{\alpha/2+1/4}} \,
K_{\alpha-1/2}\!\left(\frac{|\Delta x|}{(D_{\beta}\tau)^{1/2}}\right)\\
&\simeq& \frac{1}{2^\alpha \Gamma(\alpha)} \frac{|\Delta x|^{\alpha-1}}{(D_{\beta} \tau)^{\alpha/2}} \, \exp\left[-\frac{|\Delta
x|}{(D_{\beta}\tau)^{1/2}}\right] \! ,
\label{eqn:prop_prdc}
\end{eqnarray}
\end{subequations}
where~$\mathcal{N} = 2^{1/2-\alpha}/[\sqrt{\pi} \,\Gamma(\alpha)]$ is a normalization constant 
and $K_{\nu}(x)$ denotes the modified Bessel function of the second kind.
Asymptotically, an exponential tail emerges with a power-law correction in
$|\Delta x|$~[Eq.~\eqref{eqn:prop_prdc}] that vanishes in the case of exponentially distributed diffusion coefficients~($\alpha=1$)~\cite{noteSI}. 
By derivation, the displacement PDF depends on the similarity variable $|\Delta x|/\sqrt{\tau}$
only, reflecting the normal diffusive behavior $\Delta x^2\sim\tau$. The PDF~(\ref{eqn:prop_prdc_a}) is compared to experimental data in
Fig.~\ref{fig:msd_dist_overview}$\text{(d-f)}$. Note that his is not a fit since all
parameters in Eqs.~\eqref{eqn:prop_prdc_t} were derived from the empirical PDF
of diffusion coefficients~(Fig.~\ref{fig:gam_dist_D}). The agreement of
the displacement PDFs confirms our hypothesis that individual cargo
particles perform Brownian motion with randomly distributed diffusivities. 

\emph{Discussion.} We found that the nonequilibrium transport of microparticles in an active bath of cells displays a crossover from superdiffusion to Fickian transport with pronounced non-Gaussian displacement PDFs. While the dynamics of each particle becomes Brownian, the diffusivity varies across the ensemble~\cite{chechkin_brownian_2017}.
This variation may arise from tug-of-war between multiple cells simultaneously attached to the cargo particle.
The tug-of-war may lead to repeated unsuccessful attempts to move the cargo in a given time window.
Intermittent motion with distributed immobilization events can be described by subordination of a parent process with a waiting time PDF~\cite{magdziarz_2009_fractional,fox_2021_aging}.
If the subordinator is a gamma distribution, Brownian motion stays Fickian yet the displacements follow a Laplace PDF~\cite{madan_1990_new,madan_1998_variance}. 
If the parent process is fractional Brownian motion, a Gaussian
process with power-law correlated increments~\cite{mandelbrot_fractional_1968,khadem_stochastic_2022}, the subordination by the gamma
distribution produces a PDF with stretched tails proportional to~$\exp(-a | \Delta x |^{\delta} )$, the stretching exponent of which is $\delta=2/(1+\gamma)$~\cite{kozubowski_2006_fractional}.
The DACF derived from our experimental data at short
times indeed shows positive values beyond the rescaled time shift $\Delta/\tau=1$, reminiscent of fractional Brownian motion.
Thus, assuming fractional Laplace motion (FLM) for the cargo particle transport dynamics, superdiffusion in the short-time regime with an exponent of~$\gamma \approx 1.45$~(median of the observed exponents, cf.~Fig.~\ref{fig:msd_dist_overview}) would imply a stretching exponent of~$\delta \approx 0.82$, which is close to the experimentally inferred~$\delta \approx 0.77$ as shown in Fig.~\ref{fig:msd_dist_overview}(c). 
Assuming that the power-law correlations have a finite cutoff that reflects the observed crossover from super- to Fickian diffusion, the motion at long times would change from $\langle\Delta\mathbf{r}(\tau)^2 \rangle\simeq t^{\gamma}$ to $\simeq t$~\cite{molina_2018_crossover}, and the displacement PDF exhibits exponential tails~($\delta=1$), in line with our experimental observations. 
More refined data will be needed to connect the observed motion with potential tug-of-war immobilization events.
Furthermore, it will be of interest to discuss asymptotic Laplace displacement PDFs and FLM in a wider context of anomalous diffusion processes in the future~\cite{sabri_2020_elucidating,fox_2021_aging,munoz_2021_objective,seckler_2022_bayesian}.

Non-Gaussian statistics due to~(dynamic) heterogeneity has already been reported for acetylcholine receptors on live muscle cell membranes~\cite{he_dynamic_2016} and for cytoplasmic mRNA molecules in both {\it E.~coli} and yeast~\cite{lamp_cytoplasmic_2017}. Here, we demonstrate that anomalous effects may also arise at the level of interacting cells when collectively moving passive microobjects. Since cell-cell heterogeneity and fluctuation-dominated dynamics are ubiquitous in biological systems, our findings are relevant beyond our specific model system, for instance, when microparticles are exposed to migrating neutrophils. Generally, foreign bodies that interact with a dynamic tissue environment are key to many medical applications---oral vaccination strategies~\cite{soares_oral_2018} or the assimilation of environmental microplastics in the body~\cite{pulvirenti_effects_2022} both rely on the intestinal uptake of microparticles. Options to guide the cell-driven microtransport by chemical gradients~\cite{Nagel2018} make this process particularly attractive for drug delivery applications.

\begin{acknowledgments}

This research has been partially funded by Deutsche
Forschungsgemeinschaft~(DFG), grant 318763901--SFB1294 (R.G., L.B., T.M.,
and C.B.), 116203121~--~BE 3978/3-3 (S.S.P. and C.B.), and ME 1535/12-1 (R.M.).
We thank Maike Stange for valuable comments on the manuscript and support
regarding the experiment, as well as Kirsten Sachse for supporting lab
routines.

\end{acknowledgments}

%

\end{document}